\documentclass[preprint,pteplogo]{ptephy_v2}

\preprintnumber{arXiv:2504.09110v2}

\usepackage{amsmath,amssymb,amsthm}
\usepackage{xcolor}
\usepackage[utf8]{inputenc}
\usepackage{booktabs}
\usepackage{amsmath}
\usepackage{amsfonts}
\usepackage{braket}
\usepackage{bm}
\usepackage{hyperref}

\newcommand{\diff}{\mathrm{d}}

\begin{document}

\title{Gaussian process regression with additive periodic kernels for pairwise interaction analysis in coupled phase oscillators}

\author[1,*]{Haruma Furukawa}
\affil[1]{Graduate School of Informatics, Kyoto University, Yoshida-Honmachi, Sakyo-ku,
Kyoto, 606-8501, Japan \email{furukawa.haruma.64n@st.kyoto-u.ac.jp}}
\author[1]{Ryosuke Yoneda}
\author[1]{Daigo Fujiwara}
\author[1]{Toshio Aoyagi}

\begin{abstract}%
    Synchronization phenomena in rhythmic systems are commonly modeled using phase oscillator frameworks in which interactions are represented as sums of pairwise coupling functions. Accurate estimation of these coupling functions from observed data remains a fundamental challenge, particularly when observations are sparse or biased due to synchronization. In this paper, we propose a nonparametric Bayesian approach based on Gaussian process regression for estimating coupling functions in phase oscillator models. We model each pairwise interaction function as a periodic Gaussian process and represent the total interaction as their sum. To incorporate this structure, we introduce an additive periodic kernel that reflects both the periodicity of phase interactions and their pairwise-additive form. The proposed method is validated using coupled van der Pol oscillators, coupled FitzHugh--Nagumo oscillators, and networks of Hodgkin--Huxley-type neurons. Numerical experiments demonstrate that the proposed method provides more accurate estimates than a conventional Fourier-based Bayesian approach under sparse-data and synchronization-biased conditions. These improvements also lead to more accurate reconstruction of neuronal oscillator networks. The results illustrate how additive Gaussian-process models can incorporate prior knowledge about interaction structure through kernel design and provide a flexible framework for estimating pairwise interaction functions from oscillatory data.
\end{abstract}

\subjectindex{A55}

\maketitle

\section{Introduction}\label{sec:introduction}

Rhythm synchronization is a ubiquitous phenomenon observed in nature and engineering, in which macroscopic order emerges from the collective interactions of rhythmic units. Many such systems can be modeled as networks of nonlinear limit-cycle oscillators. When perturbations acting on each oscillator are sufficiently weak, phase reduction theory enables a substantial simplification of the dynamics by representing each oscillator solely through its phase variable while preserving the essential features of synchronization phenomena~\cite{Kuramoto1984,Winfree1980,Nakao2016}. The resulting phase description provides a mathematically tractable framework for analyzing collective behavior in large oscillator networks. In many applications, interactions are assumed to be pairwise, leading to phase oscillator models in which the phase dynamics of each oscillator is expressed as the sum of coupling functions associated with neighboring oscillators. Such models have been successfully used to describe rhythmic dynamics in a wide variety of physical, biological, and engineering systems~\cite{Ermentrout1991,Viriyopase2018,Strogatz2005}.

With the increasing availability of observational data, estimating interaction functions directly from measured dynamics has become an important problem in nonlinear science, particularly for network reconstruction. Representative approaches exploit the periodic nature of phase response curves and coupling functions through Fourier-series expansions~\cite{Galan2005,Tokuda2007,Tokuda2019,ota2014}. In particular, Tokuda et al.~\cite{Tokuda2007,Tokuda2019} developed a multiple-shooting framework for reconstructing coupling functions from oscillatory data, while Ota et al.~\cite{ota2014} introduced a linear Bayesian framework that determines the truncation order of the Fourier expansion through model-evidence maximization. More recently, Matsuki et al.~\cite{Matsuki2025} proposed a circle-map-based method for network reconstruction and demonstrated accurate connectivity inference even in synchronized regimes.

A practical difficulty in coupling-function estimation arises from the non-uniform distribution of observed phase differences. When oscillators become partially or fully synchronized, trajectories occupy only a limited region of phase space, resulting in strongly biased sampling of interaction functions. Rosenblum and Pikovsky pointed out that, in synchronized states, phase trajectories fail to explore the full phase domain, making the reconstruction of phase interactions fundamentally difficult~\cite{Rosenblum2001}. Tokuda et al. further demonstrated that the accuracy of coupling-function estimation deteriorates as synchronization develops because phase differences cease to sufficiently cover the phase space~\cite{Tokuda2007,Tokuda2019}. Motivated by this limitation, Matsuki et al.~\cite{Matsuki2025} developed a reconstruction framework specifically designed to improve inference performance in synchronized regimes. These studies suggest that synchronization-induced sampling bias constitutes a major obstacle to accurate reconstruction of interaction functions and network connectivity.

A further challenge arises from the limited availability of observational data. In many biological and experimental systems, only a small number of oscillation cycles can be recorded, resulting in sparse observations of phase interactions. Under such conditions, estimating a large number of Fourier coefficients may lead to overfitting and increased uncertainty. The problem becomes even more severe when synchronization or experimental constraints produce highly uneven sampling of phase space.

To overcome these challenges, we propose a novel approach based on Gaussian process regression, a nonparametric kernel-based method~\cite{Rasmussen2006}, for estimating coupling functions in phase oscillator models.
A key requirement in this setting is that the model preserve the additive structure of pairwise interactions, which general-purpose regression methods do not guarantee.
Our approach explicitly incorporates two fundamental properties of phase interaction functions---periodicity and pairwise additivity---into the kernel design through an additive periodic kernel. The periodic structure guarantees periodicity of the inferred coupling functions, while the additive structure enables reconstruction of individual pairwise interactions. Because GPR naturally quantifies uncertainty and regularizes function estimation through Bayesian inference, the proposed framework remains effective even when the available observations are sparse or unevenly distributed. Through numerical experiments on coupled oscillator networks, we demonstrate that the proposed method reconstructs coupling functions more accurately than a conventional Fourier-based Bayesian approach and achieves improved network reconstruction performance under challenging data conditions.

This paper is organized as follows. In Section~\ref{sec:preliminaries}, we provide an overview of coupled phase oscillator models. Section~\ref{sec:method} presents a method for estimating coupling functions using Gaussian process regression. In Section~\ref{sec:numerical}, we apply the proposed method to several systems. First, using coupled van der Pol oscillators and coupled FitzHugh--Nagumo oscillators, we demonstrate the superiority of our method over a baseline Bayesian linear regression when the dataset is small or biased. Next, we apply both methods to networks of Hodgkin--Huxley-type neurons and demonstrate that our proposed method can reconstruct the underlying network with higher accuracy. Finally, in Section~\ref{sec:conclusion}, we summarize this paper and discuss several limitations and potential future applications.

\section{Coupled phase oscillators}
\label{sec:preliminaries}
Let us consider a system comprising $N$ interacting dynamical systems:
\begin{align}\label{eq:orig_model}
    \dot{\bm{X}_i}= F_i(\bm{X}_i)+ \sum_{\substack{j=1\\j\neq i}}^{N}G_{j\to i}(\bm{X}_i,\bm{X}_j)\quad i=1,2,\cdots,N.
\end{align}
Here, $F_i$ represents the intrinsic dynamics of element $i$, and $G_{j\to i}$ represents a weak interaction from element $j$ to element $i$. We assume that each element possesses a stable limit cycle, so that it behaves as an oscillator, and that the interactions are sufficiently weak. Under these assumptions, the system can be reduced to a set of coupled phase oscillators~\cite{Kuramoto1984,Winfree1980}:
\begin{align}\label{eq:phase_oscillator}
    \dot{\phi_i} = \omega_i + \sum_{\substack{j=1\\j\neq i}}^{N} \Gamma_{ij}(\Delta \phi_{ij}), \quad \Delta \phi_{ij}:=\phi_{j}-\phi_{i}.
\end{align}
The functions $\{\Gamma_{ij}\colon \mathbb{T}\to\mathbb{R}\}_{i,j}$ are called coupling functions, where $\Gamma_{ij}$ represents the influence of oscillator $j$ on oscillator $i$ and $\mathbb{T}:=[0,2\pi)$.
Throughout this paper, the terms ``interaction function'' and ``coupling function'' are used interchangeably.
We emphasize that the coupling function depends only on the phase difference, rather than on the respective phases, $\phi_i$ and $\phi_j$.

The primary advantage of the phase-oscillator description is its substantial reduction in dimensionality. While the original system~\eqref{eq:orig_model} is typically high-dimensional, the reduced model~\eqref{eq:phase_oscillator} assigns the one-dimensional phase $\phi_i$ to each oscillator. Consequently, the analysis can focus exclusively on synchronization properties.

\section{Methods for Estimating Coupling Functions}\label{sec:method}
In this section, we present the problem formulation, the Gaussian process regression framework for estimation, and the method for phase reconstruction from observed variables.

\subsection{Time-series data setting}\label{subsec:data setting}
Consider a phase time series discretized with a time step $\delta t$.
Let the phase of oscillator $i$ at time $t=\ell\delta t$ be denoted by $\phi_i^{(\ell)}$,
and define $\bm{\phi}^{(\ell)} := (\phi_1^{(\ell)}, \cdots, \phi_N^{(\ell)}) \in \mathbb{T}^{N}$.
The entire time series is represented as $\{\bm{\phi}^{(\ell)}\}_{\ell=1}^{n_{\mathrm{data}}}$, where $n_{\mathrm{data}}$ denotes the number of data points.
In practice, the phases are not observed directly but are reconstructed from an observed scalar variable;
the reconstruction procedure is described in Section~\ref{sec:phase reconstruction}.
Throughout Sections~\ref{subsec:data setting}--\ref{subsec:optimization}, we treat the reconstructed phase time series as given.

The dataset used for estimating the dynamics consists of pairs of state variables and their temporal derivatives,
defined as $\mathcal{D}=\{(\bm{\phi}^{(\ell)}, \bm{y}^{(\ell)})\}_{\ell=1}^{n_{\mathrm{data}}}$. 
Here,
\begin{align}
\bm{y}^{(\ell)} := \left( \frac{\phi_{1}^{(\ell+1)}-\phi_{1}^{(\ell)}}{\delta t},\cdots,\frac{\phi_{N}^{(\ell+1)}-\phi_{N}^{(\ell)}}{\delta t} \right)^{\top}\in\mathbb{R}^{N}
\end{align}
is a finite-difference approximation of the phase velocity.

To relate these data to the phase model~\eqref{eq:phase_oscillator}, we make two modeling assumptions explicit.
First, the deviation of $y_i^{(\ell)}$ from the true phase velocity---originating from the finite-difference approximation, from noise acting on the underlying dynamics, and from errors in the phase reconstruction---is modeled collectively as additive Gaussian observation noise.
Second, because a constant function cannot be distinguished from the natural frequency $\omega_i$ on the right-hand side of Eq.~\eqref{eq:phase_oscillator}, we absorb $\omega_i$ into the coupling functions:
we introduce effective coupling functions $\Gamma'_{ij}$, each equal to $\Gamma_{ij}$ up to an additive constant, chosen such that $\sum_{j\neq i}\Gamma'_{ij}(\Delta\phi_{ij})=\omega_i+\sum_{j\neq i}\Gamma_{ij}(\Delta\phi_{ij})$.

Under these assumptions, the time-series data satisfy
\begin{align}
\begin{split}
    y_i^{(\ell)} &= f_i(\bm{x}_i^{(\ell)}) + \xi_i^{(\ell)}, \quad
    f_i(\bm{x}_i^{(\ell)}) = \sum_{\substack{j=1\\j\neq i}}^{N} \Gamma_{ij}'(\Delta \phi_{ij}^{(\ell)}), \\
    \bm{x}_i^{(\ell)} &= (\Delta\phi_{i1}^{(\ell)},\cdots,\Delta\phi_{i\,i-1}^{(\ell)},\Delta\phi_{i\,i+1}^{(\ell)},\cdots,\Delta\phi_{iN}^{(\ell)})^{\top} \in \mathbb{T}^{N-1},
\end{split}
\end{align}
where $\Delta\phi_{ij}^{(\ell)} := \phi_j^{(\ell)} - \phi_i^{(\ell)}$ for $i \neq j$, and $\xi_i^{(\ell)}$ is an additive noise term following $\mathcal{N}(0,\sigma_i^{2})$.

In the data-driven estimation of the coupling functions, an intrinsic uncertainty exists in the form of an additive constant. This ambiguity originates from the fact that, when only the coupled system is observed without access to the uncoupled dynamics of individual oscillators, it is impossible to disentangle how the natural frequencies $\omega_i$ are modified by the interactions from the observed data alone. Therefore a direct comparison between the estimated and the true coupling functions may exhibit a constant offset even when the estimation is otherwise accurate. Since the additive constant is not identifiable from the coupled time series alone, we compare coupling functions up to an additive constant. For visualization, the curves are shifted by an additive constant when comparing their shapes.

\subsection{Definition of the stochastic model}
Our task is to estimate the coupling functions $\{\Gamma_{ij}'\}$ from the data $\mathcal{D}$. As a direct consequence of phase reduction theory, $\Gamma_{ij}'$ is a $2\pi$-periodic function. Furthermore, we assume that $\Gamma_{ij}'$ is continuous. This assumption is empirically reasonable, and all systems used for numerical validation in later sections satisfy this condition.

To model $\Gamma_{ij}'$ nonparametrically while reflecting its smoothness, we place a Gaussian process prior on $\Gamma'_{ij}$:
\begin{align}
    \Gamma'_{ij}\sim\mathcal{GP}(m_{ij}, k_{ij}),
\end{align}
where $m_{ij}\colon\mathbb{T}\to\mathbb{R}$ is called a mean function and $k_{ij}\colon\mathbb{T}\times\mathbb{T}\to\mathbb{R}$ is called a kernel function. 

When each $\Gamma_{ij}'$ is modeled as an independent Gaussian process, their sum is also a Gaussian process,
\begin{align}\label{eq: additive GP}
    f_i\sim \mathcal{GP}(m_{i}, k_{i}),
\end{align}
with
\begin{align}\label{eq: additive GP m and k}
    m_i(\bm{x}):=\sum_{j\neq i}m_{ij}(x_j),\quad k_i(\bm{x},\bm{x}'):=\sum_{j\neq i}k_{ij}(x_j,x'_j).
\end{align}
This is referred to as an additive Gaussian process.
Furthermore, because all component processes $\{\Gamma_{ij}'\}$ are Gaussian processes, the collection consisting of the summed process $f_i$ and all individual components $\Gamma_{ij}'$ forms a jointly Gaussian process. Consequently, conditioning on the observations $\mathcal D$ yields the joint posterior distribution of $f_i$ and all $\Gamma_{ij}'$ simultaneously.
The posterior distributions of the individual coupling functions are therefore obtained directly as marginal distributions of this joint posterior.
Details regarding the mean functions and kernel functions used in this study are described in Section~\ref{subsec:cov_func}.

The posterior distribution of $f_i$ is calculated analytically, and $f_i\mid\mathcal{D}$ is again a Gaussian process. 
The distribution is given by the following closed form:
\begin{align}\label{eq: exact posterior}
    f_i\mid\mathcal{D}\sim\mathcal{GP}(\overline{m}_i,\overline{k}_i),
\end{align}
where the posterior mean function $\overline{m}_i$ and the posterior kernel function $\overline{k}_i$ are
\begin{align}\label{eq: exact mean and kernel}
    \begin{split}
        \overline{m}_i(\bm{x})&=m_i(\bm{x})+K_{i;\bm{x}X}(K_{i;XX}+\sigma_{i}^{2}I_{n_{\mathrm{data}}})^{-1}(\bm{y}_i-m_{i;X}),\\
    \overline{k}_{i}(\bm{x},\bm{x}')&=k_{i}(\bm{x},\bm{x}')-K_{i;\bm{x}X}(K_{i;XX}+\sigma_{i}^{2}I_{n_{\mathrm{data}}})^{-1}K_{i;X\bm{x}'},
    \end{split}
\end{align}

with
\begin{align*}
    [K_{i;XX}]_{mn}&=k_i(\bm{x}_{i}^{(m)},\bm{x}_{i}^{(n)}), \\
    K_{i;X\bm{x}}&=K_{i;\bm{x}X}^{\top}=(k_i(\bm{x}_{i}^{(1)},\bm{x}),\dots,k_i(\bm{x}_{i}^{(n_{\mathrm{data}})},\bm{x}))^{\top}, \\
    m_{i;X}&=(m_i(\bm{x}_i^{(1)}),\dots,m_i(\bm{x}_i^{(n_{\mathrm{data}})}))^{\top}, \\
    \bm{y}_i&=(y_i^{(1)},\dots,y_i^{(n_{\mathrm{data}})})^{\top}.
\end{align*}

The exact posterior of each individual coupling function is obtained in the same manner. From Eq.~\eqref{eq: additive GP m and k} and the prior independence of the component processes, the cross-covariance between a component and the sum is
\begin{align}\label{eq:cross_cov_data}
\operatorname{Cov}\bigl[\Gamma'_{ij}(x),\, f_i(\bm{x}')\bigr] = k_{ij}(x, x'_j), \qquad x\in\mathbb{T},\ \bm{x}'\in\mathbb{T}^{N-1},
\end{align}
because only the $j$-th component of the sum is correlated with $\Gamma'_{ij}$. Hence $\Gamma'_{ij}$ and the noisy observations $\bm{y}_i$ are jointly Gaussian, and conditioning on $\mathcal{D}$ yields
\begin{align}\label{eq:exact_component_posterior}
    \Gamma'_{ij}\mid\mathcal{D}\sim\mathcal{GP}(\overline{m}_{ij},\overline{k}_{ij}),
\end{align}
with
\begin{align}\label{eq:exact_component_mean_kernel}
    \begin{split}
        \overline{m}_{ij}(x)&=m_{ij}(x)+K_{ij;xX}(K_{i;XX}+\sigma_i^{2}I_{n_{\mathrm{data}}})^{-1}(\bm{y}_i-m_{i;X}),\\
        \overline{k}_{ij}(x,x')&=k_{ij}(x,x')-K_{ij;xX}(K_{i;XX}+\sigma_i^{2}I_{n_{\mathrm{data}}})^{-1}K_{ij;Xx'},
    \end{split}
\end{align}
where
\begin{align*}
    K_{ij;Xx}=K_{ij;xX}^{\top}=(k_{ij}(x,\Delta\phi_{ij}^{(1)}),\dots,k_{ij}(x,\Delta\phi_{ij}^{(n_{\mathrm{data}})}))^{\top}
\end{align*}
collects the covariances between $\Gamma'_{ij}(x)$ and the training points, in which only the $j$-th coordinate $\Delta\phi_{ij}^{(\ell)}$ of each input $\bm{x}_i^{(\ell)}$ appears, by virtue of Eq.~\eqref{eq:cross_cov_data}. Note that Eq.~\eqref{eq:exact_component_mean_kernel} involves the component mean function $m_{ij}$, whereas only the sum $m_i$ is specified in our model (Section~\ref{subsec:cov_func}); consequently, the posterior means of the individual components are determined only up to additive constants. This is consistent with the intrinsic ambiguity discussed in Section~\ref{subsec:data setting} and does not affect any of the analyses in this paper, where coupling functions are compared up to an additive constant.

Equations~\eqref{eq: exact posterior}--\eqref{eq:exact_component_mean_kernel} completely specify the proposed method in its idealized form: given the data, the mean function, and the kernels, the posterior distribution of every pairwise coupling function is available in closed form. Evaluating these expressions, however, requires inverting an $n_{\mathrm{data}}\times n_{\mathrm{data}}$ matrix, which is prohibitive for long time series. The remainder of this section addresses this practical issue: Section~\ref{subsec:svgp} introduces a sparse variational approximation of the posterior, and Section~\ref{subsec:optimization} describes how all adjustable parameters are determined.

\subsection{Choice of the mean functions and the kernel functions}
\label{subsec:cov_func}
We now specify the kernel function used in our Gaussian process model.
Since $f_i$ is $2\pi$-periodic in each argument, a periodic kernel is appropriate. Moreover, because the coupling function is additive (i.e., expressed as a sum of pairwise interactions), the kernel is chosen to have an additive structure, as in Eq.~\eqref{eq: additive GP m and k}.

We adopt the following periodic kernel for each component process:
\begin{align}\label{eq:add_k}
    k_{ij}(x,x')=\sigma_{ij}^{2}\exp\left\{-\frac{\sin^2[(x-x')/2]}{2\tau_{ij}^2}\right\}.
\end{align}
Here, the hyperparameter $\sigma_{ij}$ controls the amplitude of $k_{ij}$, while $\tau_{ij}$ plays the role of a length scale, determining the smoothness of $\Gamma'_{ij}$ and hence of $f_i$ along the $j$-th direction. 

We also adopt a constant function as the mean function $m_{ij}$.
However, this leads to a redundant parameterization, since the sum of constant functions is again a constant.
To avoid this redundancy, we instead treat $m_i$ as a single constant function, $m_i(\bm{x})\equiv c_i$, whose value $c_i$ is treated as a hyperparameter, without explicitly decomposing it into pairwise components.
Accordingly, wherever a component mean $m_{ij}$ appears in Eqs.~\eqref{eq:exact_component_mean_kernel} and \eqref{eq:svgp_component_mean_kernel}, any decomposition satisfying $\sum_{j\neq i}m_{ij}=m_i$ (e.g., $m_{ij}\equiv c_i/(N-1)$) may be adopted;
the choice affects the estimated components only through additive constants, which are not identifiable in any case, as discussed in Section~\ref{subsec:data setting}.

\subsection{Sparse Variational Gaussian Processes}\label{subsec:svgp}

A major computational bottleneck in Gaussian process regression is the inversion of the kernel matrix
\begin{align}
    (K_{i;XX}+\sigma_{i}^{2}I_{n_{\mathrm{data}}})^{-1},
\end{align}
which requires $O(n_{\mathrm{data}}^3)$ time and $O(n_{\mathrm{data}}^2)$ memory. To address this issue, sparse variational approximations based on inducing variables have been proposed.

In the following, we switch from the function-space representation to a finite-dimensional representation evaluated at the training inputs \(X_{i}=\{\bm{x}_{i}^{(\ell)}\}_{\ell=1}^{n_{\mathrm{data}}}\).
We introduce inducing variables
\begin{align}
    \mathbf{u}_{i}=f_i(Z_{i})
\end{align}
defined at $M (\ll n_{\mathrm{data}})$ inducing inputs $Z_{i}=\{\mathbf{z}_{i}^{(m)}\}_{m=1}^M$, $\mathbf{z}_{i}^{(m)}\in\mathbb{T}^{N-1}$, together with the latent variables
\begin{align}
    \mathbf{f}_{i}=f_{i}(X_{i}).
\end{align}
Their joint prior distribution is given by
\begin{align}
p(\mathbf{f}_{i}, \mathbf{u}_{i})=
\mathcal{N}
\left(
\begin{bmatrix}
    \bm{m}_{i;f} \\
    \bm{m}_{i;u}
\end{bmatrix},
\begin{bmatrix}
K_{i;ff} & K_{i;fu}\\
K_{i;uf} & K_{i;uu}
\end{bmatrix}
\right),
\end{align}
which yields the conditional distribution
\begin{align}\label{eq:conditional_fu}
    p(\mathbf{f}_i\mid \mathbf{u}_i)
    =
    \mathcal{N}
    \left( \bm{m}_{i;f} + 
    K_{i;fu}K_{i;uu}^{-1}(\mathbf{u}_{i}-\bm{m}_{i;u}),
    \;
    K_{i;ff}-K_{i;fu}K_{i;uu}^{-1}K_{i;uf}
    \right),
\end{align}
where
\begin{align*}
    [\bm{m}_{i;f}]_{n} &= m_i(\bm{x}_i^{(n)}), \\
    [\bm{m}_{i;u}]_{n} &= m_i(\mathbf{z}_{i}^{(n)}), \\
    [K_{i;ff}]_{mn} &= k_{i}(\bm{x}_{i}^{(m)},\bm{x}_i^{(n)}), \\
    [K_{i;fu}]_{mn} &= k_i(\bm{x}_i^{(m)},\mathbf{z}_{i}^{(n)}), \\
    [K_{i;uf}]_{mn} &= [K_{i;fu}]_{nm}, \\
    [K_{i;uu}]_{mn} &= k_i(\mathbf{z}_{i}^{(m)},\mathbf{z}_{i}^{(n)}).
\end{align*}

Classical sparse prior approximations such as FITC approximate the conditional covariance by its diagonal component~\cite{fitc}. Instead, we employ a variational approximation based on inducing variables~\cite{titsias09}.

Specifically, we use the sparse variational Gaussian process (SVGP) framework proposed by Hensman \textit{et al.}~\cite{hensman13,hensman15}. The exact posterior $p(\mathbf{f}_i,\mathbf{u}_i\mid\bm{y}_i)$ is approximated by a variational distribution of the form
\begin{align}\label{eq:variational_family}
    q(\mathbf{f}_i,\mathbf{u}_i)
    =
    p(\mathbf{f}_i\mid\mathbf{u}_i)\,q(\mathbf{u}_i),
    \qquad
    q(\mathbf{u}_i)
    =
    \mathcal{N}(\bm{\mu}_i,\mathbf{S}_i),
\end{align}
where the conditional $p(\mathbf{f}_i\mid\mathbf{u}_i)$ is fixed to the prior conditional~\eqref{eq:conditional_fu} and only $q(\mathbf{u}_i)$ is adjustable. The free parameters of the approximation are thus the variational mean $\bm{\mu}_i\in\mathbb{R}^{M}$ and the variational covariance $\mathbf{S}_i\in\mathbb{R}^{M\times M}$, the latter parameterized through its Cholesky factor to ensure positive definiteness, together with the inducing inputs $Z_i$. These variational parameters are determined, jointly with the hyperparameters, by maximizing the evidence lower bound, as described in Section~\ref{subsec:optimization}. In the remainder of this subsection, we assume that the optimized $q(\mathbf{u}_i)$ is given and describe how the posterior of each coupling function is computed from it.

The approximate posterior of each component process $\Gamma'_{ij}$ is obtained by replacing the exact conditioning on the data in Eq.~\eqref{eq:exact_component_posterior} with conditioning through the inducing variables. Since $\mathbf{u}_i=f_i(Z_i)$, Eq.~\eqref{eq:cross_cov_data} gives the cross-covariance
\begin{align}\label{eq:cross_cov_u}
    \operatorname{Cov}\bigl[\Gamma'_{ij}(x),\, [\mathbf{u}_i]_m\bigr]
    =
    k_{ij}(x, z_{i;j}^{(m)}),
\end{align}
where $z_{i;j}^{(m)}$ denotes the $j$-th component of the inducing input $\mathbf{z}_i^{(m)}$. Hence $\Gamma'_{ij}(x)$ and $\mathbf{u}_i$ are jointly Gaussian, and the conditional $p(\Gamma'_{ij}\mid\mathbf{u}_i)$ is a Gaussian process whose mean is linear in $\mathbf{u}_i$. Marginalizing $\mathbf{u}_i$ over $q(\mathbf{u}_i)$,
\begin{align}\label{eq:svgp_component_def}
    q(\Gamma'_{ij})
    :=
    \int p(\Gamma'_{ij}\mid\mathbf{u}_i)\,q(\mathbf{u}_i)\,d\mathbf{u}_i,
\end{align}
therefore yields again a Gaussian process,
\begin{align}\label{eq:svgp_component_posterior}
    \Gamma'_{ij}
    \sim
    \mathcal{GP}(\widetilde{m}_{ij},\widetilde{k}_{ij})
    \quad
    \text{under } q(\Gamma'_{ij}),
\end{align}
with
\begin{align}\label{eq:svgp_component_mean_kernel}
    \begin{split}
        \widetilde{m}_{ij}(x)
        &=
        m_{ij}(x)
        +
        K_{ij;xu}K_{i;uu}^{-1}(\bm{\mu}_i-\bm{m}_{i;u}),
        \\
        \widetilde{k}_{ij}(x,x')
        &=
        k_{ij}(x,x')
        -
        K_{ij;xu}K_{i;uu}^{-1}K_{ij;ux'}
        +
        K_{ij;xu}K_{i;uu}^{-1}\mathbf{S}_i K_{i;uu}^{-1}K_{ij;ux'},
    \end{split}
\end{align}
where
\begin{align*}
    K_{ij;ux}=K_{ij;xu}^{\top}=(k_{ij}(x,z_{i;j}^{(1)}),\dots,k_{ij}(x,z_{i;j}^{(M)}))^{\top}
\end{align*}
collects the covariances~\eqref{eq:cross_cov_u} between $\Gamma'_{ij}(x)$ and the inducing variables. The first two terms of $\widetilde{k}_{ij}$ constitute the covariance conditioned on $\mathbf{u}_i$, while the third term propagates the uncertainty of $q(\mathbf{u}_i)$ into the prediction. Equation~\eqref{eq:svgp_component_mean_kernel} has the same structure as the exact posterior~\eqref{eq:exact_component_mean_kernel}, with the $n_{\mathrm{data}}$-dimensional data statistics replaced by $M$-dimensional statistics of the inducing variables; consequently, once $q(\mathbf{u}_i)$ has been optimized, the cost of evaluating the posterior of any coupling function is independent of $n_{\mathrm{data}}$. The same substitution applied to the sum $f_i$ yields its approximate posterior $q(f_i)$, which appears in the evidence lower bound in Section~\ref{subsec:optimization}. The estimated coupling functions reported in Section~\ref{sec:numerical} are the posterior mean of Eq.~\eqref{eq:svgp_component_posterior}.

\subsection{Optimization of the variational parameters and hyperparameters}
\label{subsec:optimization}

The model for oscillator $i$ contains two groups of adjustable quantities: (i) the parameters of the variational approximation, namely $\bm{\mu}_i$, $\mathbf{S}_i$, and the inducing inputs $Z_i$; and (ii) the hyperparameters, namely the constant $c_i$ of the mean function, the kernel parameters $\{\sigma_{ij},\tau_{ij}\}_{j\neq i}$, and the observation noise standard deviation $\sigma_i$. These are the only quantities subject to optimization. All of them are determined jointly by maximizing the evidence lower bound (ELBO)~\cite{hensman13,hensman15}
\begin{align}\label{eq:elbo}
    \mathcal{L}_i
    =
    \sum_{\ell=1}^{n_{\mathrm{data}}}
    \mathbb{E}_{q(f_i^{(\ell)})}
    \left[
    \log \mathcal{N}\bigl(y_i^{(\ell)}\,\big|\, f_i^{(\ell)},\, \sigma_i^{2}\bigr)
    \right]
    -
    \mathrm{KL}
    \left(
    q(\mathbf{u}_i)\,\|\,p(\mathbf{u}_i)
    \right)
    \;\leq\;
    \log p(\bm{y}_i),
\end{align}
where $f_i^{(\ell)}:=f_i(\bm{x}_i^{(\ell)})$ denotes the $\ell$-th entry of $\mathbf{f}_i$, and $q(f_i^{(\ell)})$ is the corresponding marginal of
\begin{align}\label{eq:q_f}
    q(\mathbf{f}_i)
    =
    \int p(\mathbf{f}_i\mid\mathbf{u}_i)\,q(\mathbf{u}_i)\,d\mathbf{u}_i.
\end{align}
Both terms in Eq.~\eqref{eq:elbo} are available in closed form for the Gaussian likelihood.

The maximization of the ELBO plays a twofold role. First, for fixed hyperparameters, maximizing $\mathcal{L}_i$ with respect to $(\bm{\mu}_i,\mathbf{S}_i,Z_i)$ is equivalent to minimizing the Kullback--Leibler divergence $\mathrm{KL}\bigl(q(\mathbf{f}_i,\mathbf{u}_i)\,\|\,p(\mathbf{f}_i,\mathbf{u}_i\mid\bm{y}_i)\bigr)$; that is, it selects the best approximation to the exact posterior within the variational family~\eqref{eq:variational_family}. The resulting $\bm{\mu}_i$ and $\mathbf{S}_i$ are precisely the quantities entering the predictive distribution~\eqref{eq:svgp_component_mean_kernel}. Second, since $\mathcal{L}_i$ is a lower bound on the log marginal likelihood $\log p(\bm{y}_i)$, its maximization with respect to the hyperparameters plays the role that evidence maximization plays in exact Gaussian process regression. In other words, the ELBO is maximized not only to determine the hyperparameters but also to optimize the variational parameters $\bm{\mu}_i$ and $\mathbf{S}_i$. In our implementation, all parameters are updated simultaneously by stochastic gradient ascent using the Adam optimizer (see Appendix~\ref{app:simulation} for the settings).

Because the expected log-likelihood in Eq.~\eqref{eq:elbo} decomposes over data points, unbiased estimates of $\mathcal{L}_i$ and its gradient can be computed from mini-batches, enabling stochastic optimization. The resulting per-iteration cost is $O(BM^2+M^3)$, where $B$ is the mini-batch size, in contrast to the $O(n_{\mathrm{data}}^3)$ cost of the exact posterior~\eqref{eq: exact mean and kernel}. Finally, since each oscillator is modeled independently, the objectives $\{\mathcal{L}_i\}_{i=1}^{N}$ are optimized independently for each $i$.

In practice, $B$ is taken to be of order $M$ or larger, which is natural because the memory requirement is already $O(M^2)$; then $M^3+M^2B=O(M^2B)$, and one epoch, consisting of $n_{\mathrm{data}}/B$ steps, costs $O(M^2 n_{\mathrm{data}})$. The sparse approximation thus reduces the dependence on the data size from the cubic $O(n_{\mathrm{data}}^3)$ of the exact posterior~\eqref{eq: exact mean and kernel} to a linear one.

\subsection{Phase reconstruction}\label{sec:phase reconstruction}
The phase time series is reconstructed from the observed variable $\mathcal{O}(t)$. 
We first calculate the angle of the analytic signal:
\begin{align}
    \vartheta(t) = \arg \{\hat{\mathcal{O}}(t)+\mathrm{i}\operatorname{\mathcal{H}}[\hat{\mathcal{O}}](t)\},
\end{align}
where $\hat{\mathcal{O}}(t)$ is obtained from $\mathcal{O}(t)$ by choosing its sign so that the phase of the analytic signal increases monotonically along the limit cycle and $\operatorname{\mathcal{H}}[\hat{\mathcal{O}}]$ is the Hilbert transform of $\hat{\mathcal{O}}$.

Since $\vartheta(t)$ does not generally evolve uniformly in time, we transform it following~\cite{Kralemann2007,Kralemann2008}:
\begin{align}
    \phi(t)=2\pi\int_{0}^{\vartheta(t)}\sigma(\varphi)\mathrm{d}\varphi,
\end{align}
where $\sigma$ is the empirical distribution of $\vartheta$. 
The resulting $\phi(t)$ is regarded as the oscillator phase and used for regression. This reconstruction procedure is entirely data-driven and thus directly applicable to experimental data, although its accuracy may depend on the specific system.

\section{Numerical Simulations}
\label{sec:numerical}

\subsection{General numerical setup}
\label{subsec:procedure}
We first describe the general procedure common to all numerical experiments.
For details on parameters, difference schemes, and other aspects of numerical calculation, see Appendix~\ref{app:simulation}.

\subsubsection{Reference functions and baseline estimates}\label{sec:theory and baseline}

To assess the validity and performance of the estimation results obtained using the proposed method, we plot the theoretically derived coupling function and the baseline estimation results alongside those obtained using the proposed method.

Theoretical coupling functions are calculated by averaging the product of the phase sensitivity and the coupling term over $\mathbb{T}$. The phase sensitivity is obtained by solving the adjoint equation pertaining to the limit-cycle oscillator~\cite{Nakao2016}.

We use Bayesian linear regression as our baseline estimation method proposed in Ref.~\cite{ota2014}. This method estimates the posterior distributions of the coefficients in a Fourier-series expansion of the coupling functions. Model selection is performed to appropriately determine the truncation order of the series expansion. Details can be found in Appendix~\ref{app:model selection}.

\subsubsection{Data generation}
All systems are simulated as stochastic differential equations with additive noise. 
The data size is controlled by two parameters: $N_{\rm init}$, the number of initial conditions, and $T$, the duration of each simulation. The relationship between them and $n_{\mathrm{data}}$ and the method for determining $T$ is described in Appendix~\ref{sec:simulation time}.

\subsection{Coupled van der Pol oscillators}
\label{sec:vdp}
We first apply the proposed method to coupled van der Pol oscillators ($N=2$). In the form of Eq.~\eqref{eq:orig_model}, we define the state of oscillator $i$ as $\bm X_i=(x_i,y_i)^\top$ and the intrinsic dynamics as
\begin{align*}
    \bm{F}_i=\begin{pmatrix}
    y_i \\
    \varepsilon_i(1-x_i^2)y_i-x_i
    \end{pmatrix},
\end{align*}
where $\varepsilon_{1}=0.68$ and $\varepsilon_{2}=0.7.$ The coupling functions are defined as
\begin{align*}
    \bm{G}_{2\to1}=
    \begin{pmatrix}
        -0.01x_1(x_2-y_1)^2 \\ 
        0.03y_1(x_2-y_1)
    \end{pmatrix},\quad 
    \bm{G}_{1\to2}=
    \begin{pmatrix}
        0.005x_1\\
        -0.03y_1y_2
    \end{pmatrix}.
\end{align*}
Independent Gaussian white noises $\{\xi_{\alpha}(t)\},$ which satisfy $\langle\xi_{\alpha}(s)\xi_{\beta}(t)\rangle=0.01^{2}\delta_{\alpha,\beta}\delta(s-t)$ for $\alpha,\beta\in\{x_{1},y_{1},x_{2},y_{2}\}$, are added to all state variables. We choose $\mathcal{O}=x_i$ as the observed variable for phase reconstruction.

\begin{figure*}[htbp]
    \centering
    \includegraphics[width=0.8\textwidth]{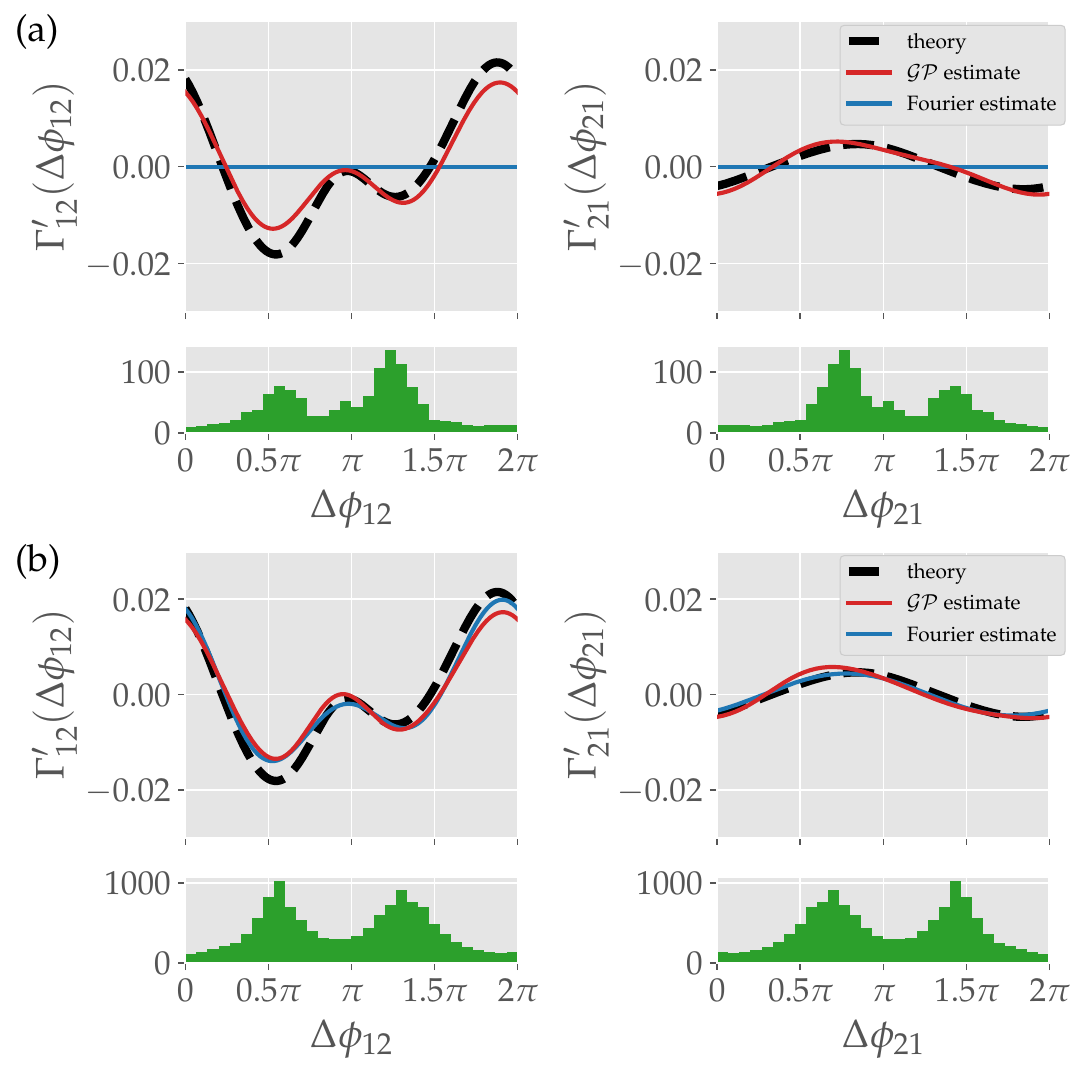}
    \caption{Estimation results for all coupling functions in coupled van der Pol oscillators, along with data histograms. The plot shows the theoretical prediction (black dashed curves), the estimation results from the proposed method (red solid curves), and the baseline estimation results (blue solid curves). (a) $(N_{\rm init},T)=(1,647)$. (b) $(N_{\rm init},T)=(10,647)$.}
    \label{fig:vdp}
\end{figure*}

The results of estimating the coupling functions for $(N_{\rm init},T)=(1,647)$ are shown in Fig.~\ref{fig:vdp}. The total number of data points is $n_{\mathrm{data}}=1254$. The corresponding data distributions are shown in Fig.~\ref{fig:vdp}.

From Fig.~\ref{fig:vdp}(a), the results obtained using our proposed method are in good agreement with the theoretical predictions, while the baseline method fails to produce accurate estimates.
This difference is due to the small amount of data; increasing the amount of data reveals that the baseline method can also provide accurate estimates [see Fig.~\ref{fig:vdp}(b)].
As mentioned in Section~\ref{sec:theory and baseline}, the baseline method performs model selection by maximizing evidence; however, when the amount of data is limited, this means that a model with a truncation order of 0 is selected.

\subsection{Coupled FitzHugh--Nagumo oscillators}
\label{sec:fn}

Next, we consider coupled FitzHugh--Nagumo oscillators~\cite{FitzHugh1961,Nagumo1962}  ($N=3$):
\begin{align*}
    \bm{X}_i &=
    \begin{pmatrix}
        v_i \\
        w_i
    \end{pmatrix}, &\quad
    \bm{F}_i&=
    \begin{pmatrix}
    v_i - \frac{1}{3}v_i^3 - w_i + 0.8 \\
    \tau_i^{-1}(v_i + a_i -0.8w_i)
    \end{pmatrix}, \\ 
    \bm{G}_{2\to1}&=
    \begin{pmatrix}
        0.01v_1v_2^3 \\ 
        0
    \end{pmatrix}, &\quad
    \bm{G}_{3\to1}&=
    \begin{pmatrix}
        0 \\
        0.005v_1w_3^2
    \end{pmatrix}, \\  
    \bm{G}_{1\to2}&=
    \begin{pmatrix}
        0 \\
        0
    \end{pmatrix}, &\quad
    \bm{G}_{3\to2}&=
    \begin{pmatrix}
        0.02(w_2-w_3) \\ 
        0
    \end{pmatrix}, \\  
    \bm{G}_{1\to3}&=
    \begin{pmatrix}
        0.01(w_3v_1-v_3w_1) \\
        0
    \end{pmatrix},&\, 
    \bm{G}_{2\to3}&=
    \begin{pmatrix}
        -0.01w_2w_3 \\
        0.005v_3(w_2-v_2)
    \end{pmatrix},
\end{align*}
where $(a_1,\tau_1^{-1})=(0.8,0.1), (a_2,\tau_2^{-1})=(0.6,0.09)$ and $(a_3,\tau_3^{-1})=(0.7,0.095)$. As in Section~\ref{sec:vdp}, Gaussian white noise with an intensity of $0.01$ is added to all variables. We choose $\mathcal{O}=v_i$ as the observed variable for phase reconstruction.

\begin{figure*}[htbp]
    \centering
    \includegraphics[width=0.95\textwidth]{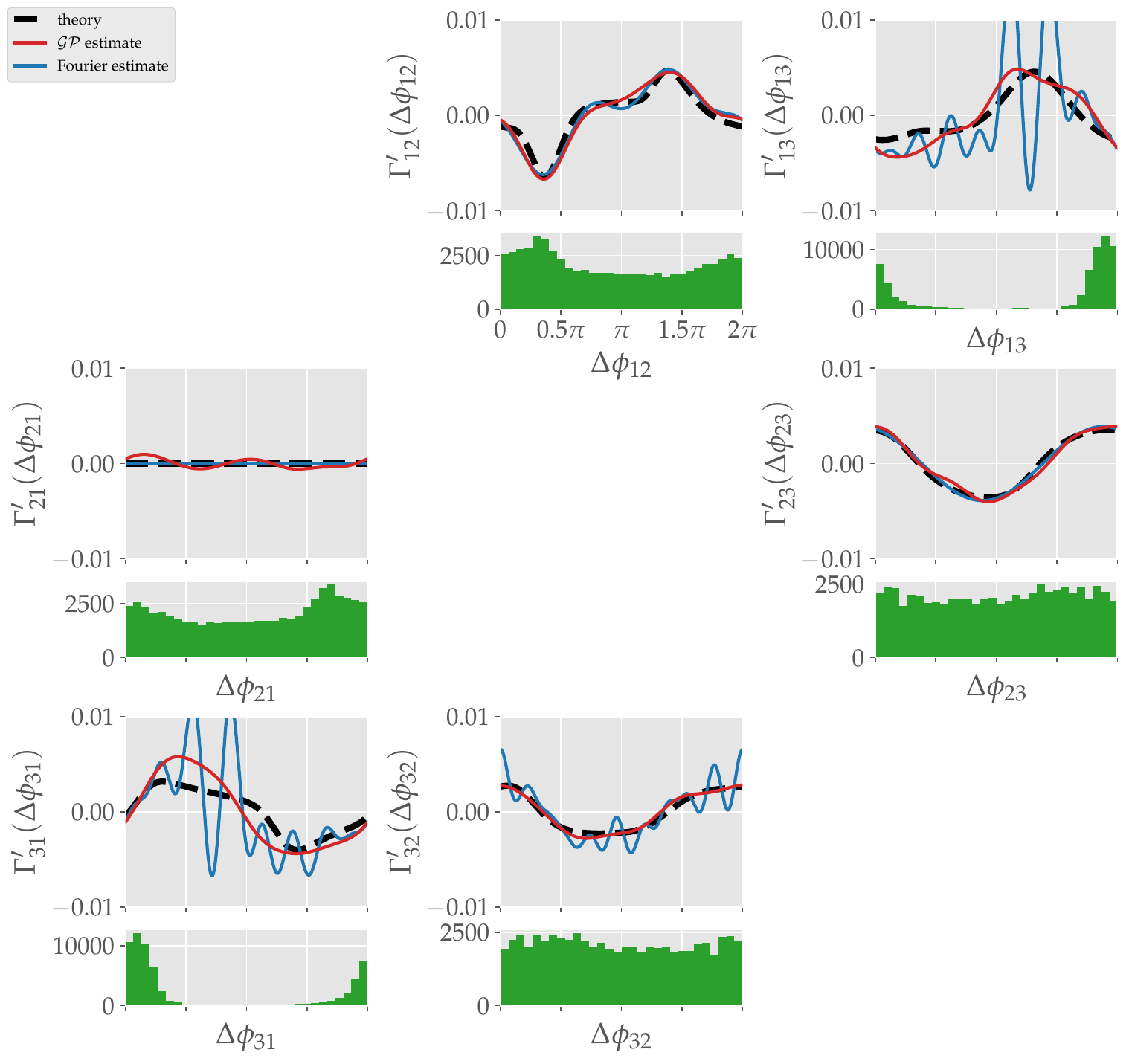}
    \caption{Estimation results for all coupling functions in the coupled FitzHugh--Nagumo oscillators, along with data histograms. The plot shows the theoretical prediction (black dashed curves), the estimation results from the proposed method (red solid curves), and the baseline estimation results (blue solid curves).}
    \label{fig:fn}
\end{figure*}

The results of estimating the coupling functions for $(N_{\rm init},T)=(2,15756)$ are shown in Fig.~\ref{fig:fn}. The total number of data points is $n_{\mathrm{data}}=62944$.

Compared to the previous section, the dataset is sufficiently large; however, there are regions in which the data distribution is strongly biased due to synchronization (see the histogram of $\Delta\phi_{13}$ or that of $\Delta\phi_{31}$ in Fig.~\ref{fig:fn}). Even in such cases, the proposed method estimates coupling functions with reasonable accuracy. However, the baseline method returns results containing high-frequency and high-amplitude components. Furthermore, even when the data are nearly uniformly distributed, the method sometimes returns results with high-frequency oscillatory components (e.g., $\Gamma_{32}$), suggesting that the model selection fails.

\subsection{Neuronal Networks}\label{sec:nn}
Finally, we consider two networks of coupled Hodgkin--Huxley-type neurons~\cite{h&h} ($N=7$ and $N=20$) and address a network reconstruction task by estimating all $N(N-1)$ pairwise coupling functions. For the $N=7$ network, five neurons are excitatory and the remaining two are inhibitory. For the $N=20$ network, sixteen neurons are excitatory and the remaining four are inhibitory. Neurons are sparsely connected via kinetic synapses [Fig.~\ref{fig:nn7}(a)]. Details of the model are summarized in Appendix~\ref{app:nn}.

Figures~\ref{fig:nn7}(c) and \ref{fig:nn20}(a) compare the coupling functions estimated by the proposed method with the theoretical coupling functions obtained via phase reduction. Overall, the estimated functions are in good agreement with the theoretical predictions, although small spurious couplings are observed for some non-connected neuron pairs.

To quantitatively evaluate the network reconstruction performance, we calculate receiver operating characteristic (ROC) curves and the corresponding area under the curve (AUC). The coupling strength from oscillator $j$ to oscillator $i$ is quantified by
\begin{align}
    \gamma_{ij}=\int_{0}^{2\pi} \widetilde{\Gamma}_{ij}(\psi)^2\,\mathrm{d}\psi,
\end{align}
which serves as a decision statistic. Here, $\widetilde{\Gamma}_{ij}$ denotes the mean-centered version of $\Gamma_{ij}$ and $\gamma_{ij}$ represents the effective magnitude of directed interaction. By varying a threshold on $\gamma_{ij}$ and comparing the inferred connections with the ground-truth network, we construct ROC curves.

The resulting ROC curves and AUC values are shown in Fig.~\ref{fig:nn7}(d) and Fig.~\ref{fig:nn20}(b). For both network sizes, the proposed method outperforms the baseline method, demonstrating its effectiveness for large-scale network reconstruction.

\begin{figure*}[htbp]
    \centering
    \includegraphics[width=0.9\textwidth]{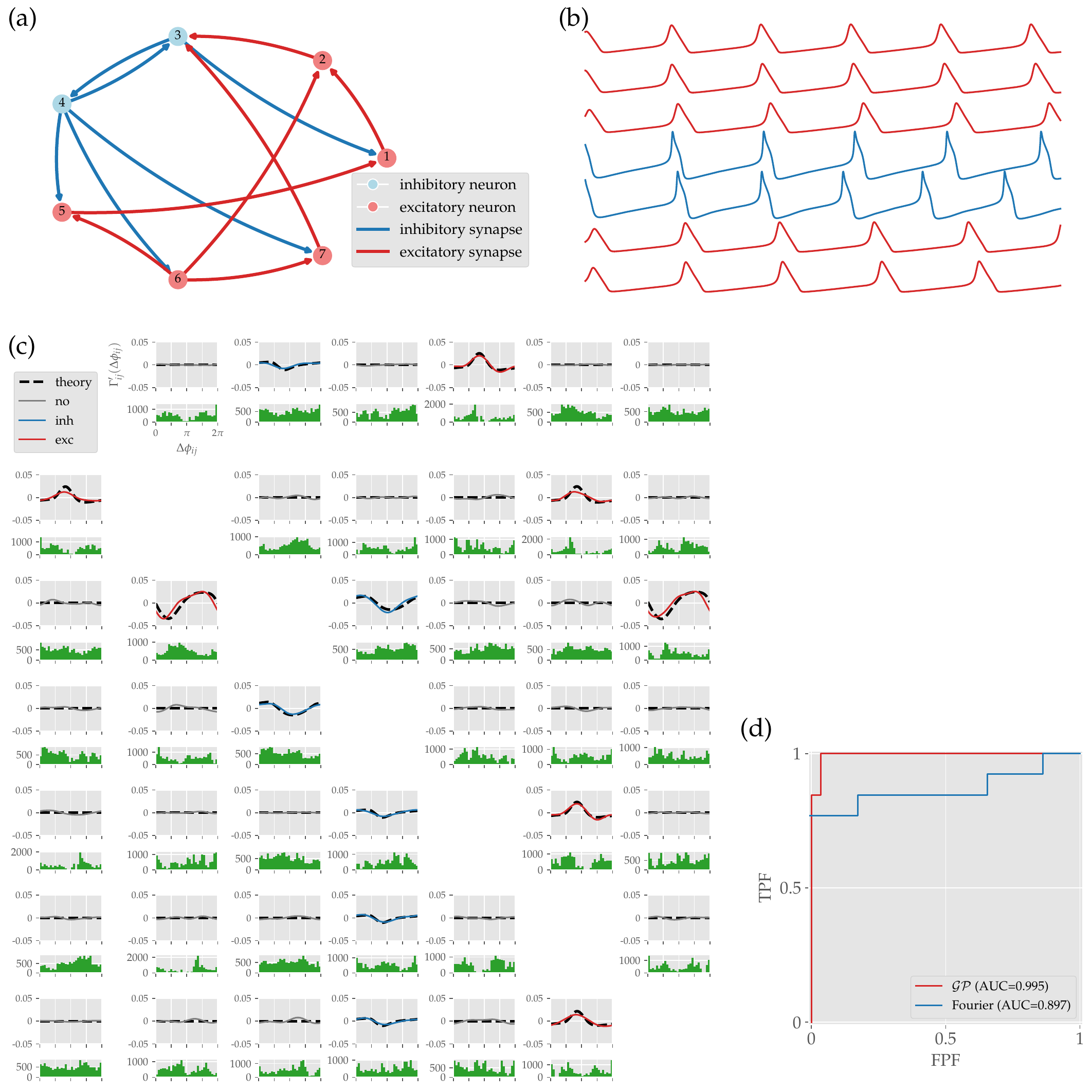}
    \caption{
    (a) Network of seven Hodgkin--Huxley-type neurons.
    (b) Time series of the membrane potentials.
    (c) Theoretical coupling functions (black dashed curves) and those estimated (solid curves: red, excitatory; blue, inhibitory; gray, no connection). Each $\Gamma_{ij}$ is shifted to have zero mean.
    (d) ROC curves and AUC values for network reconstruction. Results of the proposed method (red) and the baseline method (blue) are shown.
    }
    \label{fig:nn7}
\end{figure*}

\begin{figure*}[htbp]
    \centering
    \includegraphics[width=0.95\textwidth]{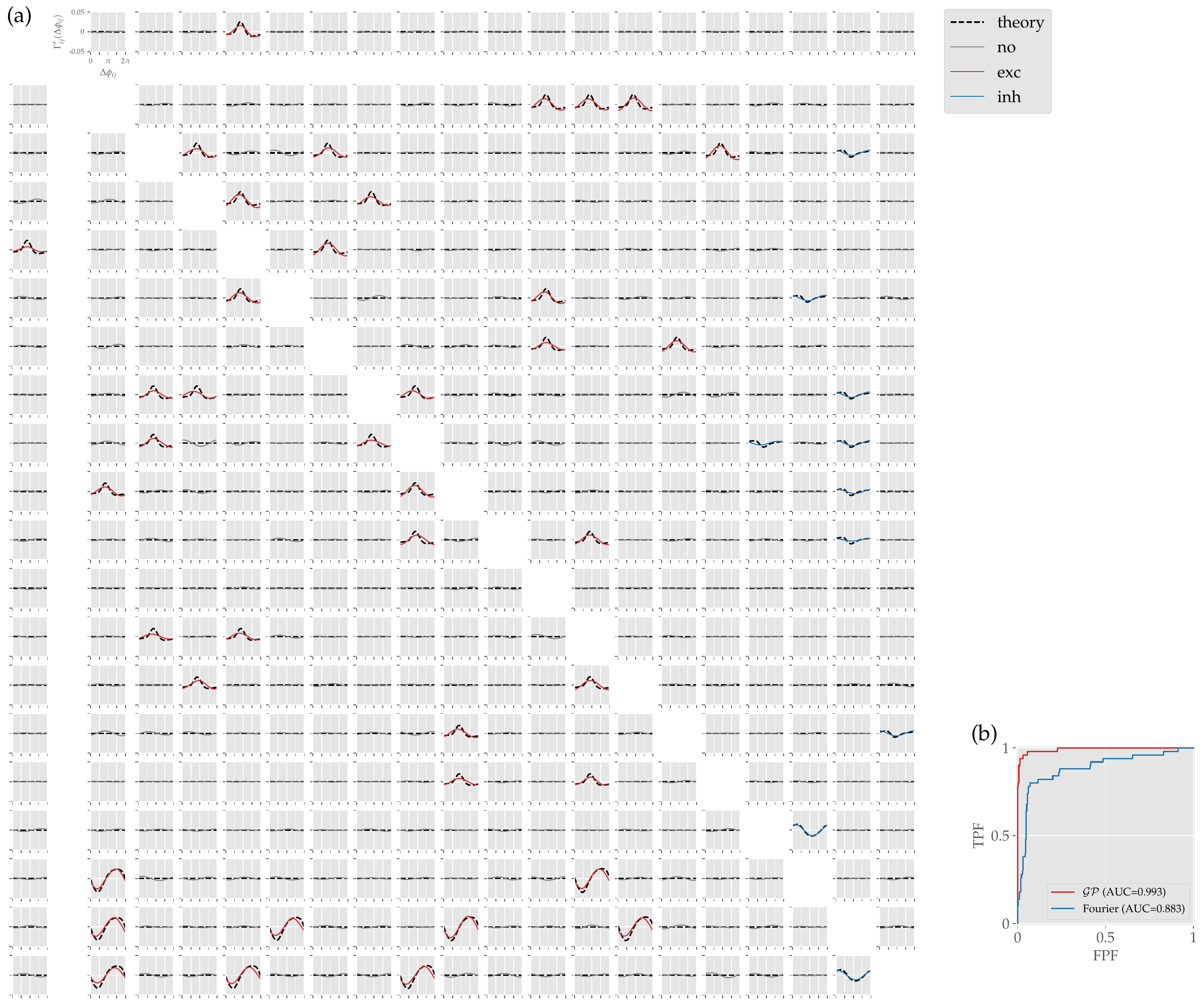}
    \caption{Reconstruction results for the $N=20$ network (16 excitatory, 4 inhibitory). 
    (a) Theoretical and estimated coupling functions. 
    (b) ROC curves comparing the proposed and baseline method.}
    \label{fig:nn20}
\end{figure*}

\section{Conclusion and Discussion}
\label{sec:conclusion}

In this paper, we proposed a Gaussian-process-based framework for estimating coupling functions in phase oscillator networks. The primary motivation of this study was to address practical challenges arising in network reconstruction from oscillatory data, particularly the presence of sparse observations and synchronization-induced sampling bias. To this end, we introduced an additive periodic kernel that explicitly incorporates the periodicity and pairwise-additive structure of phase interaction functions. Numerical experiments using coupled van der Pol oscillators and FitzHugh--Nagumo oscillators demonstrated that the proposed method provides more accurate coupling-function estimates than a conventional Bayesian Fourier approach when the available data are limited or unevenly distributed. Furthermore, experiments on Hodgkin--Huxley-type neuronal networks showed that these improvements translate into more accurate network reconstruction.

The proposed framework highlights how prior structural knowledge can be incorporated into Gaussian process models through kernel design. In particular, we considered interaction functions that can be expressed as a sum of component functions, each depending on a specific phase difference. By constructing the kernel as a corresponding sum of one-dimensional periodic kernels, both the total interaction function and the individual pairwise components can be inferred within a unified Bayesian framework. This flexibility as a nonparametric method contributes to robust inference under sparse or biased sampling conditions.

Several limitations should be noted. As discussed in Section~\ref{sec:phase reconstruction}, the estimation accuracy depends on the method used for phase reconstruction. In this study, phases are obtained via the Hilbert transform; however, more sophisticated phase estimation techniques have recently been proposed~\cite{Matsuki2023,Yawata2024,Yamamoto2025} and may further improve performance.

In addition, the computational cost grows with the network size.
Within the sparse variational framework of Section~\ref{subsec:svgp}, one gradient step of the ELBO for a single oscillator costs $O(NM(M+B) + M^{3} + M^{2}B)$, with $M$ the number of inducing points and $B$ the mini-batch size (Section~\ref{subsec:optimization}):
the first term originates from assembling the kernel matrices $K_{i;uu}$ and $K_{i;fu}$, each entry of which is a sum over the $N-1$ component kernels, while the remaining terms represent the linear algebra on $M\times M$ matrices.
Since the oscillators are estimated independently, the cost for the entire network is $O(N^{2}M(M+B) + N(M^{3} + M^{2}B))$ per step, while the data length enters only through the number of steps per epoch, $n_{\mathrm{data}}/B$.
The quadratic term reflects the fact that $N(N-1)$ pairwise functions are estimated simultaneously, and becomes dominant once $N$ grows to a value of order $M$;
the networks considered in this paper still lie in the regime where the cost grows almost linearly with $N$, whereas applications to substantially larger networks would face the $O(N^{2})$ growth.
Reducing this cost, for example by exploiting prior knowledge of network sparsity to restrict the set of candidate interactions, constitutes an important direction for future work.

From a broader perspective, the present work is situated within recent efforts to estimate nonlinear dynamical relationships from data, including deep-learning approaches~\cite{Raissi2018,Chen2018} and kernel methods~\cite{Brunton2016,Lu2021}. In contrast to such general-purpose regressors, the estimation of pairwise interaction functions requires the model to preserve their additive structure, and the present study demonstrates that this requirement can be met through kernel design.

Finally, the idea underlying the proposed additive kernel is not restricted to rhythm synchronization. In many physical, biological, and engineering systems, interactions can be represented as superpositions of pairwise contributions. Examples include gravitational interactions, electrostatic interactions, and various networked dynamical systems. Whenever the target function can be expressed as a sum of lower-dimensional component functions, an additive Gaussian-process model provides a natural framework for incorporating such structural information. From this perspective, the present work illustrates a more general strategy for combining domain-specific structural knowledge with nonparametric Bayesian inference.

\section*{Code Availability}
The source code used to generate the numerical results in this study
will be made available from the corresponding author upon reasonable
request after publication.

\section*{Acknowledgement}
This work was supported in part by the following: MEXT KAKENHI Grant Numbers JP23H04467; JSPS KAKENHI Grant Numbers JP24H00723, JP22KJ1902, JP20K20520; JST BOOST Grant Number JPMJBS2407.

\vspace{0.2cm}
\noindent

\let\doi\relax

\appendix

\section{Details of Numerical Calculations}\label{app:simulation}

\subsection{Simulation time and the data size}\label{sec:simulation time}

This section provides additional information on the data size.

We determine the simulation time $T$ using the formula $T = \lceil N_{\mathrm{cycle}} T_{\mathrm{period}} \rceil$. Here, $T_{\mathrm{period}}$ is the period of the limit cycle of oscillator 1 under unperturbed conditions, and $N_{\mathrm{cycle}}$ is an integer representing the number of cycles to be simulated.

We use the Hilbert transform for phase reconstruction. To prevent artifacts, we exclude the first and last 10 time units of the time series before performing phase reconstruction. Therefore, the total number of data points can be calculated as follows:
\begin{align*}
    n_{\mathrm{data}}=\frac{N_{\mathrm{init}}(T-20)}{\delta t},
\end{align*}
where $\delta t$ denotes the data sampling interval. Table~\ref{tab:data_parameters} shows the parameter values used for each system.

\begin{table}[tb]
\centering
\caption{Parameter values related to the data size.}
\label{tab:data_parameters}
\begin{tabular}{lccc}
\hline
Model & $T_{\mathrm{period}}$ & $N_{\mathrm{cycle}}$ & $\delta t$ \\
\hline
van der Pol oscillators & 6.462311 & 100 & 0.5 \\
FitzHugh--Nagumo oscillators & 31.510976 & 500 & 0.5 \\
Neuronal Network $(N=7)$ & 10.127506 & 5 & 0.1 \\
Neuronal Network $(N=20)$ & 10.127506 & 5 & 0.1 \\
\hline
\end{tabular}
\end{table}

\subsection{Coupled van der Pol oscillators and coupled FitzHugh--Nagumo oscillators}

The stochastic differential equations were numerically integrated in Python using the \texttt{itoint} function of the \texttt{sdeint} package. We employed the second-order Runge--Kutta scheme for It\^o SDEs (\texttt{itoSRI2}).

For SVGP, we used \texttt{gpflow.models.SVGP} in {\it GPflow}~\cite{GPflow2017}. The number of inducing variables was set to $M=100$.
Each inducing input $\mathbf{z}_{i}^{(m)}\in\mathbb{T}^{N-1}$ was initialized by drawing each of its $N-1$ components independently and uniformly from $[0,2\pi)$.
Model parameters were optimized using the Adam optimizer with the default learning rate of $10^{-3}$, for $10^{4}$ iterations with mini-batches of size $B=100$ drawn from the shuffled dataset;
the inducing inputs $Z_i$ were optimized together with all other parameters.

\subsection{Neuronal Networks}
For the network of Hodgkin--Huxley-type neurons, we employed the Euler--Maruyama scheme with time step $0.01$. For SVGP, the number of inducing variables was set to $M=50$. The optimizer settings were identical to those in Appendix~\ref{app:simulation}.2.

\section{Neuronal Networks} \label{app:nn}
This section provides a detailed explanation of the neuronal networks discussed in Section~\ref{sec:nn}.

\subsection{Hodgkin--Huxley-type Neurons}

We describe the Hodgkin--Huxley model~\cite{h&h} in detail below.
\begin{align}
\begin{split}
    C\dot{V}=&G_{\mathrm{Na}}m^{3}h(E_{\mathrm{Na}}-V)+G_{\mathrm{K}}n^{4}(E_{K}-V)+G_{\mathrm{L}}(E_{\mathrm{L}}-V)+I_{\mathrm{input}}+\xi_{V},\\
    \dot{m}=&\alpha_{m}(V)(1-m)-\beta_{m}(V)m+\xi_{m},\\
    \dot{h}=&\alpha_{h}(V)(1-h)-\beta_{h}(V)h+\xi_{h},\\
    \dot{n}=&\alpha_{n}(V)(1-n)-\beta_{n}(V)n+\xi_{n},
\end{split}
\end{align}
with the parameter values $C=1,G_{\mathrm{Na}}=120,G_{\mathrm{K}}=36,G_{\mathrm{L}}=0.3,E_{\mathrm{Na}}=50,E_{\mathrm{K}}=-77,E_{\mathrm{L}}=-54.4$.
The auxiliary functions $\alpha_{m,h,n},\beta_{m,h,n}$ are
\begin{align*}
    \alpha_{m}(V)&=\frac{0.1(V+40)}{1-\exp[(-V-40)/10]},\\
    \alpha_{h}(V)&=0.07\exp\frac{-V-65}{20},\\
    \alpha_{n}(V)&=\frac{0.01(V+55)}{1-\exp[(-V-55)/10]},\\
    \beta_{m}(V)&=4\exp\frac{-V-65}{18},\\
    \beta_{h}(V)&=\frac{1}{1+\exp[(-V-35)/10]},\\
    \beta_{n}(V)&=0.125\exp\frac{-V-65}{80}.
\end{align*}

Another neuron model we use is as follows:
\begin{align}
\begin{split}
    C\dot{V}=&G_{\mathrm{Na}}m^{3}h(E_{\mathrm{Na}}-V)+G_{\mathrm{K}}n^{2}(E_{K}-V)+G_{\mathrm{L}}(E_{\mathrm{L}}-V)+I_{\mathrm{input}}+\xi_{V},\\
    \dot{m}=&\alpha_{m}(V)(1-m)-\beta_{m}(V)m+\xi_{m},\\
    \dot{h}=&\alpha_{h}(V)(1-h)-\beta_{h}(V)h+\xi_{h},\\
    \dot{n}=&\alpha_{n}(V)(1-n)-\beta_{n}(V)n+\xi_{n},
\end{split}
\end{align}
with the parameter values $C=1,G_{\mathrm{Na}}=112,G_{\mathrm{K}}=224,G_{\mathrm{L}}=0.1,E_{\mathrm{Na}}=55,E_{\mathrm{K}}=-97,E_{\mathrm{L}}=-70.0$.
The auxiliary functions $\alpha_{m,h,n},\beta_{m,h,n}$ are
\begin{align*}
    \alpha_{m}(V)=&\frac{40(V-75)}{1-\exp[(75-V)/13.5]},\\
    \alpha_{h}(V)=&0.0035\exp\frac{-V}{24.186},\\
    \alpha_{n}(V)=&\frac{V-95}{1-\exp[(95-V)/11.8]},\\
    \beta_{m}(V)=&1.2262\exp\frac{-V}{42.248},\\
    \beta_{h}(V)=&\frac{0.017(-51.25-V)}{\exp[(-51.25-V)/5.2]-1},\\
    \beta_{n}(V)=&0.025\exp\frac{-V}{22.222}.
\end{align*}

For each cell $i$, the input current is the sum of the bias and synaptic currents,
\begin{align}
    I_{\mathrm{input},i}=I_{\mathrm{bias},i}+\sum_{j\in\mathrm{pre}_{i}}I_{\mathrm{syn},ij}.
\end{align}
Here, $\mathrm{pre}_{i}$ denotes the set of indices of the cells that send synaptic inputs to the $i$th cell. For the network with $N=7$ (Fig.~\ref{fig:nn7}), we set the bias currents as $I_{\mathrm{bias},i}=30,32,6,6.5,34,36,38$ for $i=1,2,\dots,7$, respectively. For the network with $N=20$ (Fig.~\ref{fig:nn20}), we set the bias currents as $I_{\mathrm{bias},i}=29+i$ for $i=1,2,\cdots,16$ and $I_{\mathrm{bias},i}=6+0.25(i-17)$ for $i=17,18,\cdots,20$, respectively. For both neural network models, we choose $\mathcal{O}=V$ as the observed variable for phase reconstruction.

Additionally, a weak, independent noise function $\xi_{x,i}$ is added to the membrane voltage $V_i$ and channel variables $m_i$, $h_i$ and $n_i$.
The noise follows a Gaussian white noise distribution, with $\braket{\xi_{x,i}(t)} = 0$ and $\braket{\xi_{x,i}(t)\xi_{y,j}(s)}=\sigma^2_x \delta_{xy}\delta_{ij}\delta(t-s)$, where $x,y =V, m, h, n$, and $i$ and $j$ are the cell indices.
The noise strengths used are $\sigma_V = 0.5$ and $\sigma_m = \sigma_h = \sigma_n = 5\times10^{-6}$.

\subsection{Synapse model}

The current flowing through the synapses, $I_{\mathrm{syn},ij}$, is modeled using the kinetic synapse model~\cite{Destexhe1994}, where it is represented as
\begin{align}
    I_{\mathrm{syn},ij}=G_{ij}r_{ij}(t)[V_i(t) - E_{ij}].
\end{align}
The fraction of bound receptor proteins is represented by $r_ij$, and its dynamics are
described by the following equation:
\begin{align}
    \frac{\diff r_{ij}}{\diff t} = \alpha_{ij}T_{ij}(1-r_{ij})-\beta_{ij}r_{ij},
\end{align}
where $T_{ij}$ is the concentration of neurotransmitters, which is set to 1 when a spike is emitted by the presynaptic cell and resets to 0 after 1 millisecond. 
The constants $\alpha_{ij}$ and $\beta_{ij}$ govern the kinetics of $r_{ij}$, while $E_{ij}$ is the reversal potential and $G_{ij}$ is the synaptic conductance.
The values used for excitatory and inhibitory synapses are $(\alpha_{ij}, \beta_{ij}, E_{ij}, G_{ij}) = (1.1, 0.67, 0, 0.1)$ and $(9.8, 0.2, -75, 0.08)$ respectively.

\section{Model Selection for the Fourier-Based Baseline Method} \label{app:model selection}

As a baseline method, we employed the Bayesian Fourier series approach proposed in Ref.~\cite{ota2014}. In this framework, each coupling function is represented by a truncated Fourier series, and the truncation order is determined through Bayesian model selection. Since higher truncation orders correspond to more complex statistical models, model comparison is performed based on the model evidence, which can be calculated analytically within the Bayesian formulation.

For a network of $N$ oscillators, there are $N(N-1)$ pairwise coupling functions. Let $M_{j\to i}$ denote the truncation order assigned to the coupling function from oscillator $j$ to oscillator $i$. A statistical model is therefore specified by the collection of truncation orders $\{M_{j\to i}\}_{i\neq j}$.

\subsection{Exhaustive Model Selection}

For the van der Pol oscillator system ($N=2$) and the FitzHugh--Nagumo oscillator system ($N=3$), exhaustive model selection was feasible.

For each coupling function, the truncation order was allowed to take integer values from $0$ to $10$. Here, $M_{j\to i}=0$ corresponds to a constant coupling function. All possible combinations of truncation orders were enumerated and evaluated. The model with the largest evidence was selected and used for subsequent analyses.

\subsection{Greedy Model Selection for Large Networks}

For the neuronal network examples ($N=7$ and $N=20$), exhaustive model selection becomes computationally impractical because of the combinatorial growth of the model space. Therefore, we adopted a greedy search strategy.

The procedure is summarized as follows.

\begin{enumerate}
\item Initialize truncation orders as $M_{j\to i}=5$ for all $i\neq j$.

\item Randomly select one coupling function and decrease its truncation order by one, thereby generating a candidate model of lower complexity.

\item Re-estimate the model parameters and calculate the corresponding evidence.

\item If the evidence of the candidate model exceeds that of the current model, accept the candidate model; otherwise, retain the current model.

\item Repeat steps (2)--(4) until no further update is accepted.

\end{enumerate}

This procedure substantially reduces the computational cost of model selection. However, because only local modifications of the model are explored, the resulting model is not guaranteed to maximize the evidence globally and may correspond to a local optimum.


\begin{thebibliography}{99}
\bibitem{Kuramoto1984}
Y. Kuramoto,
Springer Berlin, Heidelberg (1984).\\
\doi{https://doi.org/10.1007/978-3-642-69689-3}

\bibitem{Winfree1980}
A. T. Winfree,
Springer New York (1980).\\
\doi{https://doi.org/10.1007/978-1-4757-3484-3}

\bibitem{Nakao2016}
H. Nakao,
Contemp. Phys., \textbf{57}(2), 188--214 (2016).\\
\doi{https://doi.org/10.1080/00107514.2015.1094987}

\bibitem{Viriyopase2018}
A. Viriyopase, R.-M. Memmesheimer, and S. Gielen,
Phys. Rev. E, \textbf{98}, 022217, (2018).\\
\doi{https://doi.org/10.1103/PhysRevE.98.022217}

\bibitem{Ermentrout1991}
G. B. Ermentrout,
J. Math. Biol., \textbf{29}(6):571--585 (1991).\\
\doi{https://doi.org/10.1007/BF00164052}

\bibitem{Strogatz2005}
S. H. Strogatz, D. M. Abrams, A. McRobie, B. Eckhardt and E. Ott,
Nature \textbf{438}, 43--44 (2005). \\
\doi{https://doi.org/10.1038/438043a}

\bibitem{Galan2005}
R. F. Gal\'an, G. B. Ermentrout, and N.~N. Urban,
Phys. Rev. Lett., \textbf{94}:158101 (2005).\\
\doi{https://doi.org/10.1103/PhysRevLett.94.158101}

\bibitem{Tokuda2007}
I. T. Tokuda, S. Jain, I. Z. Kiss, and J. L. Hudson,
Phys. Rev. Lett., \textbf{99}, 064101 (2007).\\
\doi{https://doi.org/10.1103/PhysRevLett.99.064101}

\bibitem{Tokuda2019}
I. T. Tokuda, Z. Levnajic and K. Ishimura,
Phil. Trans. R. Soc. A \textbf{377}, no. 2160 (2019).\\
\doi{https://doi.org/10.1098/rsta.2019.0015.}

\bibitem{ota2014}
K. Ota and T. Aoyagi, [arXiv:1405.4126].\\
\doi{https://doi.org/10.48550/arXiv.1405.4126}

\bibitem{Matsuki2025}
A. Matsuki, H. Kori and R. Kobayashi,
PLOS Complex Systems \textbf{2}(9): e0000063 (2025). \\
\doi{https://doi.org/10.1371/journal.pcsy.0000063}

\bibitem{Rosenblum2001}
M. G. Rosenblum and A. S. Pikovsky,
Phys. Rev. E \textbf{64}, 045202 (2001).\\
\doi{https://doi.org/10.1103/PhysRevE.64.045202.}

\bibitem{Raissi2018}
M. Raissi, P. Perdikaris, and G. E. Karniadakis, [arXiv:1801.01236]. \\
\doi{https://doi.org/10.48550/arXiv.1801.01236}

\bibitem{Chen2018}
R.~T.~Q.~Chen, Y.~Rubanova, J.~Bettencourt and D.~K.~Duvenaud,
NeurIPS 31 (2018).

\bibitem{Brunton2016}
S. L. Brunton, J. L. Proctor, and J. N. Kutz, PNAS, \textbf{113}(15):3932--3937 (2016).

\bibitem{Lu2021}
F. Lu, M. Maggioni, and S. Tang,
J. Mach. Learn. Res., \textbf{22}(32):1--67 (2021).

\bibitem{Rasmussen2006}
C.~E. Rasmussen and C. K. I. Williams,
The MIT Press (2006).\\
\doi{https://doi.org/10.7551/mitpress/3206.001.0001}

\bibitem{fitc}
E. Snelson and Z. Ghahramani,
NeurIPS (2005).


\bibitem{titsias09}
M. Titsias,
Proc. Mach. Learn. Res. pages 567--574 (2009).

\bibitem{hensman13}
J. Hensman, N. Fusi, and N. D. Lawrence,
Proceedings of the Twenty-Ninth Conference on Uncertainty in Artificial Intelligence, 282--290 (2013).

\bibitem{hensman15}
J. Hensman, A. Matthews, and Z. Ghahramani,
Proceedings of the Eighteenth International Conference on Artificial Intelligence and Statistics, vol. 38 of PMLR, pages 351--360 (2015).

\bibitem{Kralemann2007}
B. Kralemann, L. Cimponeriu, M. Rosenblum, A. Pikovsky, and R. Mrowka,
Phys. Rev. E, \textbf{76}, 055201 (2007).\\
\doi{https://doi.org/10.1103/PhysRevE.76.055201}

\bibitem{Kralemann2008}
B. Kralemann, L. Cimponeriu, M. Rosenblum, A. Pikovsky, and R. Mrowka,
Phys. Rev. E, \textbf{77}, 066205 (2008).\\
\doi{https://doi.org/10.1103/PhysRevE.77.066205}

\bibitem{FitzHugh1961}
R. FitzHugh,
Biophys. J., \textbf{1}(6), 445--466 (1961).\\
\doi{https://doi.org/10.1016/S0006-3495(61)86902-6}

\bibitem{Nagumo1962}
J.~Nagumo, S.~Arimoto, and S.~Yoshizawa,
Proceedings of the IRE, \textbf{50}(10), 2061--2070 (1962).\\
\doi{https://doi.org/10.1109/JRPROC.1962.288235}

\bibitem{h&h}
A.~L. Hodgkin and A.~F. Huxley,
J. Physiol., \textbf{117}(4), 500--544 (1952).\\
\doi{https://doi.org/10.1113/jphysiol.1952.sp004764}

\bibitem{Matsuki2023}
A. Matsuki, H. Kori and R. Kobayashi,
Sci Rep \textbf{13}, 3535 (2023). \\
\doi{https://doi.org/10.1038/s41598-023-30405-5}


\bibitem{Yawata2024}
K. Yawata, K. Fukami, K. Taira and H. Nakao,
Chaos \textbf{34} (6): 063111 (2024). \\
\doi{https://doi.org/10.1063/5.0205718}

\bibitem{Yamamoto2025}
T. Yamamoto, H. Nakao and R. Kobayashi, 
Chaos, Solitons Fractals, Vol. 191, 115913, ISSN 0960-0779 (2025). \\
\doi{https://doi.org/10.1016/j.chaos.2024.115913}

\bibitem{GPflow2017}
A. G. de~G. Matthews, M. van der Wilk, T. Nickson, K. Fujii, A. Boukouvalas, P. {Le{\'o}n-Villagr{\'a}}, Z. Ghahramani, and J. Hensman,
J. Mach. Learn. Res., \textbf{18}(40), 1--6 (2017).

\bibitem{Destexhe1994}
A.~Destexhe, Z.~F. Mainen and T.~J. Sejnowski,
Neural Comput., \textbf{6}(1), 14--18, 01 (1994). \\
\doi{https://doi.org/10.1162/neco.1994.6.1.14}

\end{thebibliography}
\end{document}